\documentclass{article}
\usepackage{spconf,amsmath,graphicx}
\usepackage{xcolor}
\usepackage{amsfonts}

\usepackage{multirow}
\usepackage{algorithm}
\usepackage{algorithmicx}
\usepackage{algpseudocode}
\usepackage{amsmath}

\usepackage{verbatim}

\title{Acoustic anomaly  Detection via Latent Regularized Gaussian Mixture Generative Adversarial Networks}
\name{Chengwei Chen\textsuperscript{1},  Pan Chen\textsuperscript{2},  Lingyu Yang\textsuperscript{1},  Jinyuan Mo\textsuperscript{1},  Haichuan Song\textsuperscript{1},   \quad Yuan Xie\textsuperscript{1}$^{\dagger}$,,  Lizhuang Ma\textsuperscript{1} \thanks{${\dagger}$ Corresponding author. }}

\name{Chengwei Chen\textsuperscript{1}, Pan Chen\textsuperscript{2}, Haichuan Song\textsuperscript{1}, Yiqing Tao\textsuperscript{1}, Yuan Xie\textsuperscript{1}$^{\dagger}$, Shouhong Ding\textsuperscript{3}, Lizhuang Ma\textsuperscript{1} \thanks{${\dagger}$ Corresponding author. }}

\address{$^{1}$ East China Normal University \\
	$^{2}$ Shanghai Jiao Tong University }
\begin{document}
%
\maketitle
\begin{abstract}

Acoustic anomaly detection aims at distinguishing abnormal acoustic signals from the normal ones. It suffers from the class imbalance issue and the lacking in the abnormal instances. In addition, collecting all kinds of abnormal or unknown samples for training purpose is impractical and time-consuming. In this paper, a novel Gaussian Mixture Generative Adversarial Network (GMGAN) is proposed under semi-supervised learning framework, in which the underlying structure of training data is not only captured in spectrogram reconstruction space, but also can be further restricted in the space of latent representation in a discriminant manner. Experiments show that our model has clear superiority over previous methods, and achieves the state-of-the-art results on DCASE dataset.
\end{abstract}
\begin{keywords}
	Acoustic anomaly detection, Semi-supervised learning, Generative Adversarial Networks, Gaussian Mixture Model, Energy estimation
\end{keywords}
\section{Introduction}
Acoustic anomaly detection task is to find the abnormal samples in audios. Since audio data has typical time series characteristics, it is commonly processed by using Recurrent Neural Network (RNN) techniques \cite{marchi2017deep} \cite{o2016recurrent}. However, RNN such as Long Short Term Memory networks (LSTM) \cite{hochreiter1997long} has some limitations in memory step size and parallel training, especially when the input data is a long sequence. To solve this problem, convolutional autoregressive architecture \cite{oord2016wavenet} was proposed. This series of models are highly parallelizable in the training phase, meaning that computation made more tractable due to effective resource utilization.

Several previous work formulate acoustic anomaly detection task as the acoustic event classification problem by using the Gaussian Mixture Model (GMM) and the Hidden Markov Model (HMM) \cite{ntalampiras2011probabilistic} to detect various types of sounds. However, it is impractical to reconstruct all possible kinds of abnormal event audios. Some propose the methods rely on machine learning algorithms which are usually trained with abnormal samples directly and extract the abnormal features. Nevertheless, to take all types of abnormal samples into account is impossible. In this paper, on the contrary, only the normal data, which is so much easier to collect, will be used for training. One sample is regarded as abnormal if it does not satisfy normal features. \emph{Then the main challenge becomes extracting the characteristic of normal acoustic signals only and exploit them to detect signals that deviate from the normality.} Several studies exploit the autoencoder based on Long Short-Term Memory units \cite{bao2017deep} trained on the normal samples only. During the testing period, it is expected to get higher image reconstruction error on abnormal spectrogram not existing in the training process. The autoencoder approaches \cite{principi2017acoustic} \cite{lu2017unsupervised} \cite{oh2018residual} outperform GMM and HMM approaches. Actually, minimizing the image reconstruction error is not original intension of the task. Our aim is to exploit more separable latent representation features between the normalities and the anomalies.

Motivated by the above limitations in previous methods, we propose a novel semi-supervised learning framework for acoustic anomaly detection, consisting of the compression network and the estimation network. For compression network, compared with conventional GAN style architecture, the auxiliary encoder and the latent regularizer are adopted to minimize the distance between the bottleneck features of original input data and encoded latent feature of the generated data, resulting in a more concentrated distribution for target class. The main function of the estimation network is to execute density estimation. Given the low dimension representation of input data from compression network, instead of using decoupled two-stage training and the standard Expectation-Maximization (EM) algorithm \cite{huber2011robust}, the estimation network estimates the parameters of GMM and evaluates the energy for samples. The estimation network helps the compression network escape from less attractive local optima and further reduce reconstruction errors.

\section{Proposed Method}

\begin{figure*}[h]
	\centering
	\includegraphics[width=0.85\linewidth]{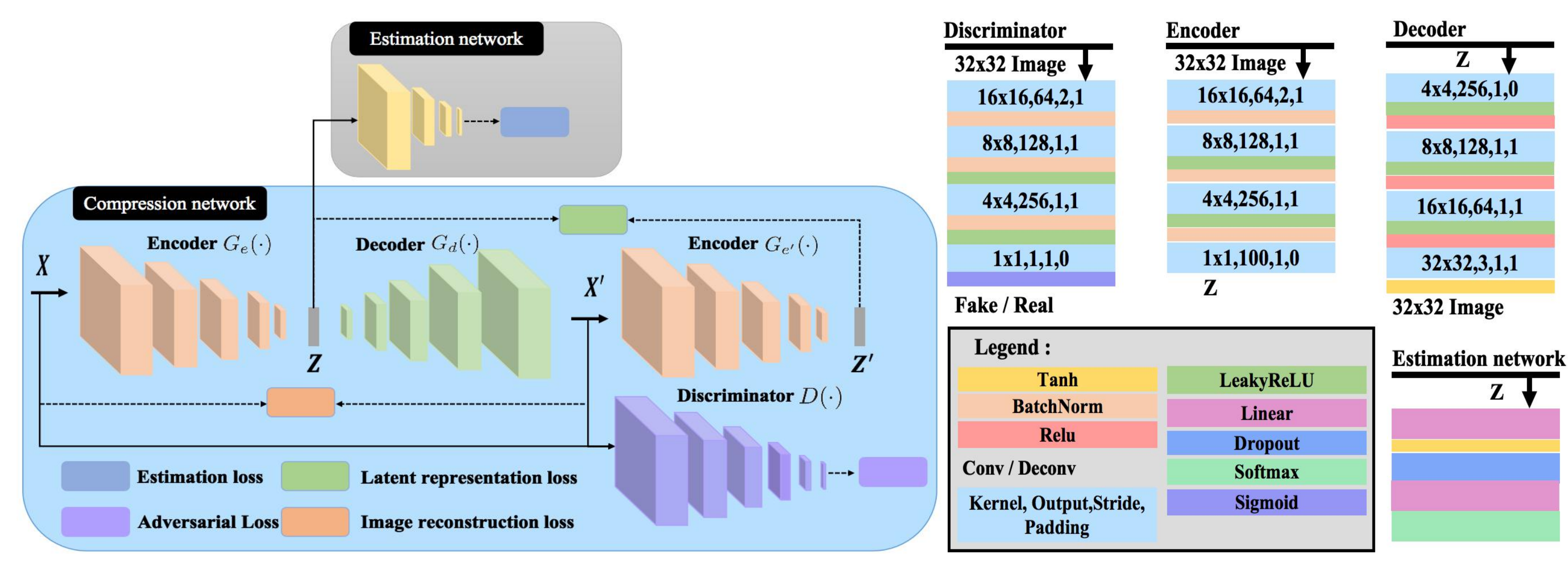}
	\caption{Our framework consists of compression network and estimation network. The architecture layers of each component are specified on the right.}
	\label{achieture}
\end{figure*}

In this section, we present more details of the proposed framework, then describe each term in loss function.

\subsection{Network Architecture}
 Inspired by the anomaly detection methods \cite{zenati2018efficient,schlegl2017unsupervised,li2018anomaly,akcay2018ganomaly}, proposed architecture is shown in Fig. \ref{achieture}, consists of two subnetworks: compression network and estimation network.

\subsubsection{Compression network}

The compression network consists the discriminator and a generative sub-network with an auxiliary encoder. The generative sub-network, including an encoder ($G_e(\cdot)$) and a decoder ($G_d(\cdot)$), attempts to reconstruct the original input sample to fool the discriminator, while the discriminator tries to distinguish original images from generated images. The model captures the real characteristic of normal samples by minimizing the distance between original input image ($\mathbf{x}$) and generated image ($\mathbf{x'}$).

The main function of the auxiliary encoder ($G_{e'}(\cdot)$) is to minimize the distance between the bottleneck features ($\mathbf{z}$) of original input image and encoded latent feature ($\mathbf{z'}$) of the generated image.
The reason why we employ an auxiliary encoder in our framework can be concluded as follows: (1) The auxiliary encoder could obtain the high-quality reconstruction of latent feature only for the normal samples. (2) Since the latent regularizer might incur the distribution distortion in latent feature space, the feature representation $\mathbf{z'}$ can be regarded as the anchor to prevent $\mathbf{z}$ from drifting.

Avoid being fooled by the generative sub-network, the discriminator learns the real concept in the target class. In addition, the discriminator also helps the generative sub-network to get more robust and stable parameters during the training period. This part of parameters could make normal and abnormal samples more separable.

\subsubsection{Estimation network}

In \cite{zong2018deep} and \cite{zhai2016deep}, Gaussian Mixture Model is greatly leveraged over the learned latent space to deal with density estimation tasks for input data with network structure. Instead of using decoupled two-stage training and the standard Expectation-Maximization (EM) algorithm \cite{huber2011robust}, we employ the estimation network in our framework to execute density estimation. In the process of training, given the low dimension representation of input data from compression network, the estimation network estimates the parameters of GMM and evaluates the energy for samples. In order to achieve the above, a multi-layer neural network regarded as the estimation network predicts the mixture membership for each sample.

\subsection{Overall Loss Function}

We define a loss function in Eqn. (\ref{loss-function}) including four components, $i.e.$, the image reconstruction loss $\mathcal{L}_{irec}$, the adversarial loss $\mathcal{L}_{adv}$, the latent representation loss $\mathcal{L}_{zrec}$ and the estimation loss $\mathcal{L}_{es}:$

\begin{equation}\label{loss-function}
\mathcal{L}=w_{i} \mathcal{L}_{irec} + w_{a} \mathcal{L}_{a d v} +w_{z} \mathcal{L}_{zrec}+w_{e} \mathcal{L}_{es}
\end{equation}
where $w_{i}$, $w_{a}$, $w_{z}$, and $w_{e}$ are the weighting parameters balancing the impact of individual item to the overall object function.

\textbf{Adversarial Loss: }GAN was first used by Goodfellow \cite{goodfellow2014generative}, which consists of a generator and a discriminator having two competing objectives:
the discriminator $D$ is learned to distinguish generated images from original input images, while the generator $G$ is trained to produce realistic images to fool $D$.  The training objective of GAN can be formulated as :

\begin{equation}
\label{eq:2}
\begin{aligned}
\mathcal{L}_{a d v}= \min _{G} \max _{D} (E_{\boldsymbol{x} \sim p_{\mathbf{x}}}[\log (D(\mathbf{x}))] \\+E_{\boldsymbol{x} \sim p_{\mathbf{x}}}[\log (1-D(G(\mathbf{x})))]).
\end{aligned}
\end{equation}

\textbf{Image reconstruction loss: } Image reconstruction loss measures the pixel-wise reconstruction error between original input image and generated image, which helps the generator to sufficiently capture the input sample distribution for normal samples. The training objective is shown below:
\begin{equation}
\label{eq:3}
\mathcal{L}_{irec}=\mathbb{E}_{x \sim p_{\mathbf{x}}}\|x-G(x)\|_{1}
\end{equation}

\textbf{Latent representation loss: }With the adversarial loss and image reconstruction loss defined above, the model can generate the realistic image. In additional to these objectives, the latent feature of generated image and that of the original input image are also expected to be consistent, which could help the generator to capture the latent feature for common examples. Hence, we consider to add a constraint by minimizing the distance between latent feature of input images $G_{e}(x)$ from generator and encoded latent feature of generated image from auxiliary encoder $G_{e^{\prime}}(x^{\prime})$ as follows.


\begin{equation}
\label{eq:4}
\mathcal{L}_{zrec}=\mathbb{E}_{x \sim p_{\mathbf{X}}}\left\|G_{e}(x)-G_{e^{\prime}}(x^{\prime})\right\|_{2}
\end{equation}

\textbf{Estimation loss: }Given the latent representation $z$ from compression network and an integer $K$ as the number of mixture components. The estimation network is a multi-layer network  followed by a softmax layer, which makes mixture-component membership prediction as follows:
$$
{\hat\gamma}=\operatorname{softmax}(M L N\left(\mathbf{z} ; \theta_{m}\right)),
$$
where $\hat\gamma$ is a \(K\)-dimensional vector for the soft mixture-component
membership prediction, which  is the output of the estimation network parameterized by  $\theta_{m}$. \(\hat\gamma_{k}\) denotes the probability of the input sample belongs to the
\(k^{\text {th } }\) distribution.  Given a batch of N samples and their membership prediction, $\forall 1 \leq k \leq K$
we can further estimate the parameters in GMM as follows:

\begin{equation}
	\begin{aligned} \hat{\alpha}_{k}=& \frac{1}{n} \sum_{i=1}^{n} \hat \gamma_{i k} ; \hat{u}_{k}=\frac{\sum_{i=1}^{n} \hat\gamma_{i k} z_{i}}{\sum_{i=1}^{n} \hat\gamma_{i k}}, \\ \hat{\Sigma}_{k}=& \frac{\sum_{i=1}^{n} \hat\gamma_{i k}\left(z_{i}-\hat{u}_{k}\right)\left(z_{i}-\hat{u}_{k}\right)^{T}}{\sum_{i=1}^{n} \hat\gamma_{i k}} ,\end{aligned}
\end{equation}

where \(\hat{\alpha}_{k}\), \(\hat{u}_{k}\),  \(\hat{\Sigma}_{k}\)  are the \(k^{t h}\) component mixture weights,
, the \(k^{t h}\) component mixture mean and the \(k^{t h}\) component mixture covariance. \(\hat\gamma_{i k}\) is the
membership prediction of the \(i^{t h}\) input sample \(z_{i}\). n is the number
of input samples.
With the GMM parameters, the energy function of an input
sample can be formulated as follows:

\begin{equation}
	E(\mathbf{z})=-\log \left(\sum_{k=1}^{K} \hat{\phi}_{k} \frac{\exp \left(-\frac{1}{2}\left(\mathbf{z}-\hat{\mu}_{k}\right)^{T} \hat{\mathbf{\Sigma}}_{k}^{-1}\left(\mathbf{z}-\hat{\mu}_{k}\right)\right)}{\sqrt{\left|2 \pi \Sigma_{k}\right|}}\right).
\end{equation}

The objective function of estimation network is to minimize
following equation,

\begin{equation}
	\mathcal{L}_{es}=\lambda_{1} \sum_{i=1}^{N} E\left(z_{i}\right)+\lambda_{2} \sum_{k=1}^{K} \sum_{j=1}^{d} \frac{1}{\hat{\Sigma}_{j j}^{k}},	
\end{equation}
where the first term is the sum of sample energy. The second term is the regularizer, which penalizes small values on the diagonal entries in covariance matrices  $ \Sigma_{k}, k=1, \ldots, K $. The purpose of this measure is to avoid the singularity problem in GMM.  $ \lambda_{1}$ and $\lambda_{2}$ are the meta parameters in estimation network.

\section{Experiments}

This section contains the description of datasets used for evaluation. Besides, some experiments present the effectiveness of the proposed method and the high performances that we obtained.

\subsection{Experimental Setting}

\begin{figure}[h]
	\centering
	\includegraphics[width=0.8\linewidth,height=0.2\linewidth]{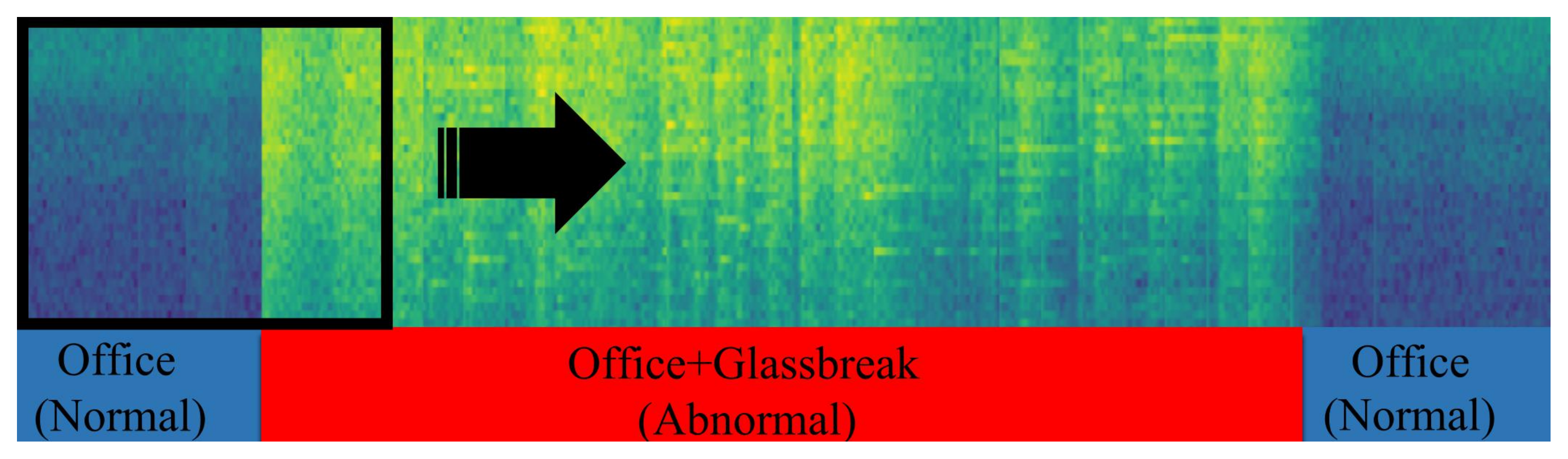}
	\caption{Spectrogram of one second sequence containing office background audio and glass break audio. }
	\label{spc}
\end{figure}
\textbf{Datasets: }We extract 15 datasets from the 2017 DCASE Challenge Task 2 dataset \cite{mesaros2017dcase}, as it is done in \cite{rushe2019anomaly}. Three different abnormal event exists in this dataset ({\it i.e.,} gunshot, babycry and glassbreak). All of these abnormal event audios are artificially mixed with background audios respectively which includes 15 different kinds of environmental settings ({\it i.e.,}home, bus, and train). Fig. \ref{spc} shows the office environment audio mixed with galss-break audio. All acoustic signals are converted to the spectrogram data, which is used as input data for proposed framework.

According to the work \cite{akcay2018ganomaly}, CIFAR10 and MNIST dataset are used to construct a toy experiment to illustrate our superiority to other state-of-the-art abnormal detectors. One of classes would be regarded as an normal class, while the rest ones belong to the anomaly class. In total, we get ten sets for MNIST dataset and CIFAR10 dataset.

\textbf{Evaluation Measures: }In testing process, the latent representation loss of each given spectrogram is regarded as an abnormal score. We compute the Area Under the ROC Curve (AUC) to measure the performance of proposed method.

\textbf{Implementation Details: } We implement our approach in PyTorch by optimizing the weighted loss $\mathcal{L}$ (defined in Eq. (\ref{loss-function}))  with the weight values $w_{i}=1$, $w_{a}=5$, $w_{z}=1$, and $w_{e}=0.05$, which are empirically chosen to yield optimum results.

\subsection{Toy  experiment}

\begin{figure}[h]
	\centering
	\includegraphics[width=\linewidth]{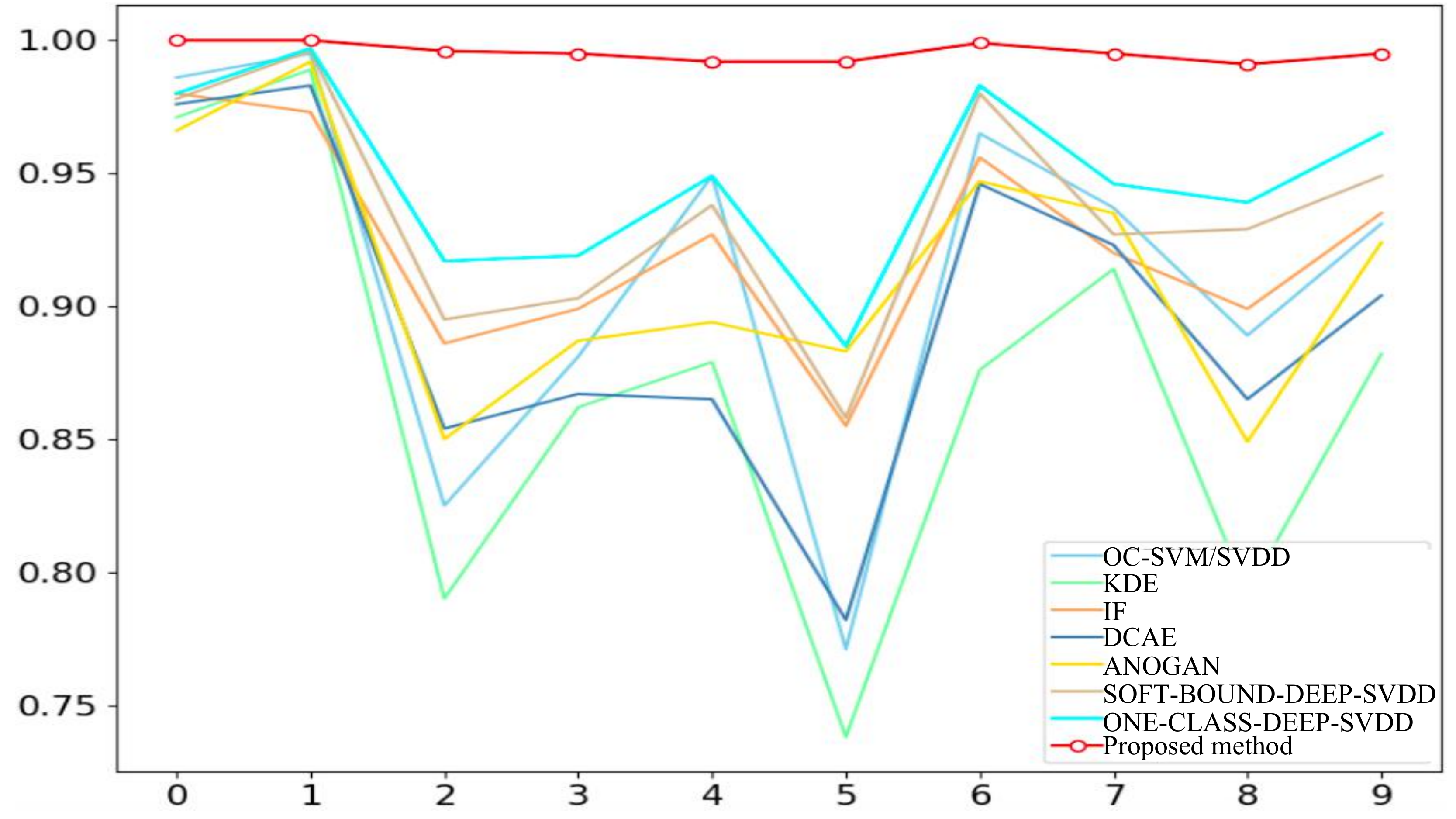}
	\caption{Digit and class designated as normal class in MNIST dataset}
	\label{minist_cifar10}
\end{figure}

In this subsection, we observe the clear superiority of our approach over previous abnormal detection models (OC-SVM/SVDD \cite{chen2001one}, KDE \cite{parzen1962estimation}, IF \cite{liu2008isolation}, DCAE\cite{masci2011stacked}, ANOGAN \cite{schlegl2017unsupervised}, DEEP-SVDD \cite{ruff2018deep}), as it is shown in Fig. \ref{minist_cifar10}. For each digit chosen as normal class in MNIST dataset, the proposed method achieves the best performance in term of AUC value, which could be illustrated by using red curves. In the right of the Fig. \ref{minist_cifar10}, we also present the comparison result in CIFAR10 dataset.

\subsection{Ablation Study}
\begin{table}[]
	\scriptsize
	\caption{The effect of different loss compositions is evaluated in five dataset experiments.}
	\label{ablationloss}
	\begin{tabular}{|l|c|c|c|c|c|}
		\hline
		&\multicolumn{5}{|c|}{Scenes  [AUC]}  \\
		
		\hline
		Loss composition&forest &store&cafe&\begin{tabular}[c]{@{}l@{}}residential \\ area \end{tabular}&train Í
	\\
		\hline
		$\mathcal{L}_{irec}$+$\mathcal{L}_{a d v}$&0.73&0.75 &0.68&0.69&0.89\\
		\hline
		$\mathcal{L}_{irec}$+$\mathcal{L}_{a d v}$+$\mathcal{L}_{zrec}$&0.75&0.82&0.74&0.76&0.90\\
		\hline
		${L}_{irec}$+$\mathcal{L}_{a d v}$+$\mathcal{L}_{zrec}$+$\mathcal{L}_{ou}$&0.77&0.83&0.76&0.78&0.92\\
		\hline
	\end{tabular}
\end{table}

\begin{figure}[h]
	\centering
	\includegraphics[width=1\linewidth]{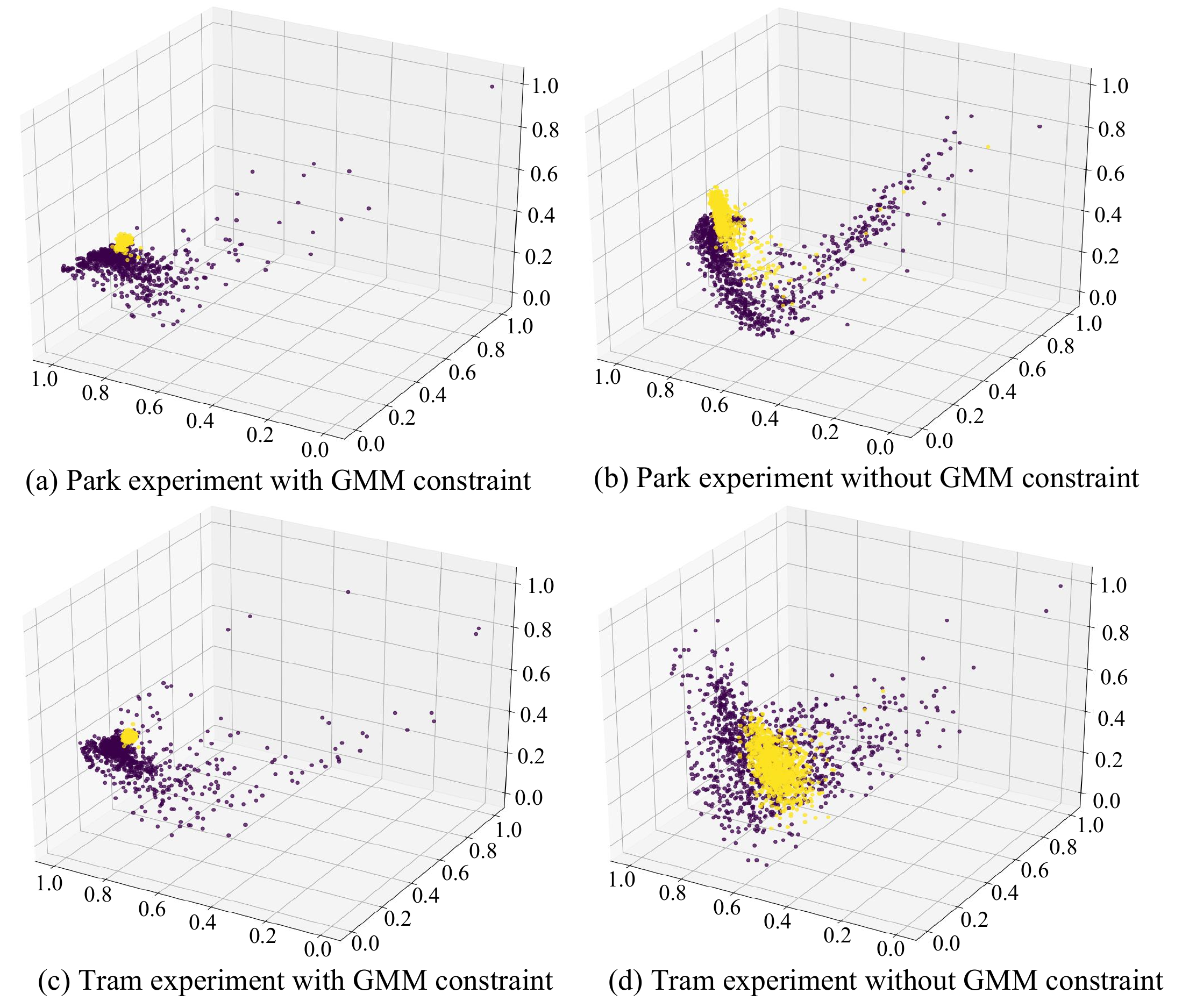}
	\caption{Proposed method learned features in 3-dimensional space  }
	\label{gmmconstraint}
\end{figure}

Since all loss functions are presented in subsection 2.2, giving the ablation study to the different combinations of loss functions is necessary. As illustrated in Tab. \ref{ablationloss}, five background-event experiments are done in the three kinds of loss combinations respectively. Each loss function we designed in the proposed framework can improve the performance.

To further explain the necessity of estimation constraint, Fig. \ref{gmmconstraint} presents the visualization of the latent feature, where the proposed method with and without  estimation constraint are compared on park and tram experiment setting. To sum up, comparison result can indicate the following: (1) Compared with the model without estimation constraint, the normal samples (yellow dots) are more concentrated with estimation constraint. (2) The anomalous samples can be more separated from normal samples (yellow dots).

\subsection{Comparison with State-of-the-art Methods}

\begin{table}[]
	\scriptsize
		\centering
	\caption{AUC scores for all methods on each dataset.}
	\label{tab:intra-dataset-loss}
	\begin{tabular}{|l|c|c|c|c|}
		\hline
		Scene&CAE \cite{oh2018residual}& WaveNet \cite{rushe2019anomaly}
	&Proposed method \\
		\hline
		beach&0.69&0.72&\textbf{0.80}\\
		\hline
		bus&0.79&0.83&\textbf{0.89}\\
		\hline
		cafe/restaurant&0.69&\textbf{0.76}&\textbf{0.76}\\
		\hline
		car&0.79&0.82&\textbf{0.93}\\
		\hline
		city center&0.75&0.82&\textbf{0.83}\\
		\hline
		forest path&0.65&0.72&\textbf{0.77}\\
		\hline
		grocery store&0.71&0.77&\textbf{0.83}\\
		\hline
		home&\textbf{0.69}&\textbf{0.69}&\textbf{0.69}\\
		\hline
		library&0.59&0.67&\textbf{0.85}\\
		\hline
		metro station&0.74&0.79&\textbf{0.81}\\
		\hline
		office&0.78&0.78&\textbf{0.80}\\
		\hline
		park&0.70&0.80&\textbf{0.89}\\
		\hline
		residential area&0.73&\textbf{0.78}&\textbf{0.78}\\
		\hline
		train&0.82&0.84&\textbf{0.92}\\
		\hline
		tram&0.80&0.87&\textbf{0.94}\\
		\hline
	\end{tabular}
\end{table}

The performance of three models across the 15 datasets is shown in Table 2. We find that the proposed model consistently outperforms the other models in almost all datasets, with a tie in the home, cafe/restaurant and residential area scenarios.

\section{Conclusion}
To capture the real characteristic of normal samples and obtain more separable latent representation features between the normalies and the anomalies, this paper exploits the technique of acoustic anomaly detection to learn an regularized latent space by using adversarial training under semi-supervised learning framework. Extensive experiments have been conducted on several datasets, showing high performance of the proposed method and the benefit of using latent regularizer from estimation network.

\bibliographystyle{IEEEbib}
\bibliography{refs}

\end{document}